\documentclass[epj]{svjour}
\input epsf

\newcounter{append}
\newcommand{\bc}{\begin{center}}
\newcommand{\ec}{\end{center}}
\newcommand{\be}{\begin{equation}}
\newcommand{\ee}{\end{equation}}
\newcommand{\ba}{\begin{array}}
\newcommand{\ea}{\end{array}}
\newcommand{\beqn}{\begin{eqnarray}}
\newcommand{\eeqn}{\end{eqnarray}}

\begin{document}

\title{Off equilibrium dynamics in disordered quantum spin chain}

\author{St\'ephane Abriet and Dragi Karevski }
\institute{Laboratoire de Physique des Mat\'eriaux, UMR CNRS No. 7556, Universit\'e Henri
Poincar\'e (Nancy 1), B.P. 239,\\ F-54506 Vand\oe uvre l\`es Nancy cedex,
France}

\date{July 1, 2002}

\abstract
{We study the  non-equilibrium time evolution of the average transverse magnetisation and
end-to-end correlation functions of the random Ising quantum chain. Starting with fully
magnetised states, either in the $x$ or $z$ direction, we compute numerically the average
quantities. They show similar behaviour to the homogeneous chain, that is an algebraic decay
in time toward a stationary state. During the time evolution, the spatial correlations, measured from one end
to the other of the chain, are building up and finally at long time they
reach a size-dependent constant depending on the
distance from criticality. Analytical arguments are given which support the numerical results.
}

\authorrunning{S. Abriet and D. Karevski }
\titlerunning{Off equilibrium dynamics in disordered quantum spin chain}


\maketitle

\maketitle

\section{Introduction}
Non-equilibrium properties of quantum spin chains at very low or even zero temperature have been the
subject of several investigations through decades. Some precursor studies were done by Niemejer~\cite{niemejer}
and Tjion~\cite{tjion} at the end of the sixties. They have shown that the relaxation is algebraic
instead of exponential as
predicted by Terwiel and Mazur using a weak coupling limit~\cite{terwiel}.
More recently, several authors have considered the relaxation
of inhomogeneously magnetised initial states and obtained analytical results for various classes of
free fermionic spin chains~\cite{berim1,racz,trimper,berim2,karevski}. It was pointed out in ref.~\cite{berim2}
that at some special wave vectors, characterizing
the inhomogeneity of the initial state, a slowing down of the relaxation can occur for free fermionic
models that have a gapped excitation spectrum. In the context of aging phenomena, two-time non-equilibrium
functions were considered in ref.~\cite{trimper,igloi1} for the Ising and XX quantum chains.
Aging occurs for intermediate times in such systems. All these out of equilibrium studies were done for
systems without any disorder although alternating and dimerized models were also considered
previously~\cite{berim2}.
In this work we study the influence of random spin couplings on the non-equilibrium relaxation properties.
For that purpose we consider the paradigmatic random Ising quantum chain.

The paper is organized as follows: in section~2 we present the random Ising quantum chain and discuss some of
its main equilibrium properties. After the canonical diagonalization, we express the time-dependent operators
we are interested in. In the next section, we first give the general expressions for the non-equilibrium
quantities we are considering, namely transverse magnetisation and end-to-end correlation functions for two
different initial states and give the numerical results and analytical analysis of this quantities.
The last section summarizes and discusses our results.
\section{Random Ising quantum chain}
The Hamiltonian of the random Ising quantum chain is given by
\begin{equation}
H=-\sum_{k}J_{k}\sigma_{k}^{x}\sigma_{k+1}^{x}-\sum_{k}h_{k}\sigma_{k}^{z}
\label{h1}
\end{equation}
where the $\sigma$s are the Pauli matrices and $J_{k}$ and $h_{k}$ are random independently
distributed couplings and transverse fields. The system possesses a quantum phase transition at
zero temperature controlled by the parameter
\begin{equation}
\delta=[\ln h]-[\ln J]\; ,
\end{equation}
where $[.]$ means an average over the disorder distribution.
The quantum control parameter $\delta$ separates the paramagnetic phase ($\delta>0$) from the ferromagnetic
one ($\delta<0$). Exactly at $\delta=0$ the system is at the critical point.
This system has been studied intensively by numerical and analytical means.
In particular, using a RG procedure
initialy introduced by Ma, Dasgupta and Hu~\cite{ma}, Fisher has obtained asymptotically
exact results (confirmed by numerical works)
on the static properties and equal time correlations in the vicinity of the critical point~\cite{fisher}.
In addition to Fisher's results, new numerical and analytical results have been obtained by various methods
and mappings away from the critical point~\cite{young,igloi2,igloi3,igloi4}, in the so called Griffiths phase~\cite{griffiths,macoy}.
In particular, the behaviour of the singular
quantities in the Griffiths phase are all characterized by a single dynamic exponent $z(\delta)$,
which is continuously varying with the control parameter $\delta$~\cite{igloi3,igloi4}. More recently,
Igl\'oi has extended
the exact RG treatment of Fisher into the Griffiths phase on both sides of the critical
point~\cite{igloi5,igloi6}.

The Hamiltonian (\ref{h1}), with free boundary conditions and $L$ sites, is readily diagonalized
after a Jordan-Wigner mapping~\cite{jordan}
\begin{eqnarray}
A_{n}=\prod_{i=1}^{n-1}(-\sigma_{i}^{z})\sigma_{n}^{x}\nonumber\\
B_{n}=i\prod_{i=1}^{n-1}(-\sigma_{i}^{z})\sigma_{n}^{y}
\end{eqnarray}
and a canonical transformation in terms of Fermi operators $\eta$ and $\eta^{+}$~\cite{lieb}:
\begin{equation}
H=\sum_{q}\epsilon_{q}\eta^{+}_{q}\eta_{q}+E_{0}
\end{equation}
where the excitation energies $\epsilon_{q}$ are the positive solutions of the $2L\times 2L$ eigenvalue
problem
\begin{equation}
{\bf T}V_{q}=\epsilon_{q}V_{q}
\end{equation}
where $\bf T$ is the tridiagonal matrix
\begin{equation}
{\bf T}=\left(\begin{array}{ccccccc}
0&h_{1}&&&&&\\
h_{1}&0&J_{1}&&&&\\
&J_{1}&0&h_{2}&&&\\
&&h_{2}&0&\ddots&&\\
&&&\ddots&\ddots& J_{L-1}&\\
&&&&J_{L-1}&0&h_{L}\\
&&&&&h_{L}&0
\end{array}\right)\; .
\end{equation}
The $2L$ components eigenvectors $V_{q}$ are usually split into two $L$ components eigenvectors $\phi_{q}$
and $\psi_{q}$, such that $V_{q}(2k-1)=-\phi_{q}(k)$ and $V_{q}(2k)=\psi_{q}(k)$.

Stationary and time-dependent expectation values of spin operators are expressed easily in terms of the
Fermi operators $\eta_{q},\eta^{+}_{q}$. Working in direct space, it is more convenient to expand these quantities
in terms of the operators $A$ and $B$ related to the diagonal fermions through
\begin{eqnarray}
A_{n}=\sum_{q}\phi_{q}(n)\left(\eta_{q}^{+}+\eta_{q}\right)\nonumber\\
B_{n}=\sum_{q}\psi_{q}(n)\left(\eta_{q}^{+}-\eta_{q}\right)\; .
\end{eqnarray}
These operators satisfy the anticommuting algebra $\{A_{n},A_{m}\}=2\delta_{n,m}$,
$\{B_{n},B_{m}\}=-2\delta_{n,m}$ and $\{A_{n},B_{m}\}=0$. From the time evolution of the diagonal
Fermi operators, $\eta_{q}^{+}(t)=e^{iHt}\eta_{q}^{+}e^{-iHt}=\eta_{q}^{+}e^{i\epsilon_{q}t}$,
one obtains the time evolution of the $A$s and $B$s:
\begin{equation}
A_{n}(t)=\sum_{m}\langle A_{n}A_{m}\rangle_{t}A_{m}+\langle A_{n}B_{m}\rangle_{t}B_{m}
\label{a1}
\end{equation}
and
\begin{equation}
B_{n}(t)=\sum_{m}\langle B_{n}A_{m}\rangle_{t}A_{m}+\langle B_{n}B_{m}\rangle_{t}B_{m}
\label{b1}
\end{equation}
where the basic contractions $\langle ..\rangle_{t}$, proportional to the time-dependent anticommutators,
are given by~\cite{igloi1,karevski}
\begin{equation}
\langle A_{n}A_{m}\rangle_{t}=\sum_{q}\phi_{q}(n)\phi_{q}(m)\cos(\epsilon_{q}t)
\end{equation}
\begin{equation}
\langle B_{n}B_{m}\rangle_{t}=\sum_{q}\psi_{q}(n)\psi_{q}(m)\cos(\epsilon_{q}t)
\end{equation}
and
\begin{equation}
\langle A_{n}B_{m}\rangle_{t}=\langle B_{m}A_{n}\rangle_{t}=i\sum_{q}\phi_{q}(n)\psi_{q}(m)\sin(\epsilon_{q}t)\; .
\end{equation}
For the homogeneous chain at the critical point, these contractions are simply expressed in terms of integer
Bessel functions~\cite{igloi1,karevski}. In the disordered situation, we evaluate them numerically.

Utilising the previous expressions, one finds for
the time evolution of the $\sigma^{z}_{l}=B_{l}A_{l}$ Pauli operator at site $l$
\begin{eqnarray}
\sigma_{l}^{z}(t)=\sum_{k}\big(\langle B_{l}B_{k}\rangle_{t}\langle A_{l}A_{k}\rangle_{t}-
\langle B_{l}A_{k}\rangle_{t}\langle A_{l}B_{k}\rangle_{t}\big)\sigma^{z}_{k}\nonumber\\
+\sum_{i\neq j}\big(\langle B_{l}B_{i}\rangle_{t}\langle A_{l}A_{j}\rangle_{t}-
\langle B_{l}A_{j}\rangle_{t}\langle A_{l}B_{i}\rangle_{t}\big)B_{i}A_{j}\nonumber\\
+\sum_{i\neq j}\langle B_{l}A_{i}\rangle_{t}\langle A_{l}A_{j}\rangle_{t}A_{i}A_{j}
+\sum_{i\neq j}\langle B_{l}B_{i}\rangle_{t}\langle A_{l}B_{j}\rangle_{t}B_{i}B_{j}\; .\nonumber\\
\label{s1}
\end{eqnarray}
Contrary to the operators $\sigma_{l}^{z}$, the operators of the form $B_{l}A_{m}$ are string operators
in terms of the Pauli matrices. For example we have for the terms $B_{l}A_{m}$:
\begin{equation}
B_{l}A_{m\ge l+1}=(-1)^{l+m+1}\sigma_{l}^{x}\left(\prod_{j=l+1}^{m-1}\sigma_{j}^{z}\right)\sigma_{m}^{x}
\label{ba1}
\end{equation}
so that in a general state it is hard to evaluate the expectations $\langle \Psi | B_{l}A_{m}|\Psi\rangle$.
In the same way, the end-to-end correlation operator $\sigma_{1}^{x}(t)\sigma_{L}^{x}(t)$, which gives
insight of the development of the correlations in the chain during the time evolution, is given by
\begin{equation}
\sigma_{1}^{x}(t)\sigma_{L}^{x}(t)=A_{1}(t)B_{L}(t){\cal Q}
\label{ssq}
\end{equation}
with
\begin{equation}
{\cal Q}=\prod_{j=1}^{L} \left(-\sigma_{j}^{z}\right)
\end{equation}
the charge operator which is a conserved quantity since $[H,{\cal Q}]=0$.
Inserting the developments (\ref{a1}) and (\ref{b1}) into the expression (\ref{ssq}),
one obtains the final form of the end-to-end correlation operator.

\section{Off equilibrium dynamics}
\subsection{Expectation values}
In what follows, we are interested in the non-equilibrium relaxation properties of the random Ising quantum
chain. The time evolution of the system is governed by the Schr\"odinger equation and is given formally by
the unitary evolution
\begin{equation}
|\Psi(t)\rangle= \exp(-iHt)|\Psi\rangle
\end{equation}
since energy is conserved. The initial states we consider in this work are eigenstates of the Pauli matrices, that
is either $|x\rangle$ or $|z\rangle$ such that $\sigma_{l}^{x,z}|x,z\rangle=|x,z\rangle$. These states
are easily accessible experimentally by applying a strong magnetic field in the desired direction.
The choice of such initial states is not only
motivated by their physical accessibility but as well because they permit close analytical formulae for the
magnetisation profile and end-to-end correlations~\cite{berim1,igloi1}. Consider first the $|z\rangle$ state.
In this state, the $\sigma^{x,y}$ operators act as flip operators and from formulae (\ref{s1}) and (\ref{ba1})
the non-vanishing expectations are only those of the $\sigma^{z}$ terms, such that finally we obtain
\begin{eqnarray}
m^{z}_{z}(l,t)\equiv\langle z|\sigma_{l}^{z}(t)|z\rangle&=&\nonumber\\
\sum_{k}\big(\langle B_{l}B_{k}\rangle_{t}\langle A_{l}A_{k}\rangle_{t}&-&
\langle B_{l}A_{k}\rangle_{t}\langle A_{l}B_{k}\rangle_{t}\big)\; .
\end{eqnarray}
Utilising ${\cal Q}|z\rangle=(-1)^{L}|z\rangle$, the end-to-end correlation function in the $z$ state for a chain
of even site number is given by
\begin{eqnarray}
C_{L}^{z}(t)\equiv\langle z|\sigma^{x}_{1}(t)\sigma_{L}^{x}(t)|z\rangle&=&\nonumber\\
\sum_{k}\big(\langle A_{1}B_{k}\rangle_{t}\langle B_{L}A_{k}\rangle_{t}&-&
\langle A_{1}A_{k}\rangle_{t}\langle B_{L}B_{k}\rangle_{t}\big)\; .
\end{eqnarray}
If the initial state is the $|x\rangle$ state, the role of the flip operators are played by the
$\sigma^{y,z}$ operators. Then, the contributing terms into (\ref{s1}) are only the $B_{l}A_{l+1}=\sigma_{l}^{x}
\sigma_{l+1}^{x}$, whose expectation value is $1$ in $|x\rangle$. This leads to
\begin{eqnarray}
m^{z}_{x}(l,t)\equiv\langle x|\sigma_{l}^{z}(t)|x\rangle&=&\nonumber\\
\sum_{k}\big(\langle B_{l}B_{k}\rangle_{t}\langle A_{l}A_{k+1}\rangle_{t}&-&
\langle B_{l}A_{k+1}\rangle_{t}\langle A_{l}B_{k}\rangle_{t}\big)\; .
\end{eqnarray}
The connected end-to-end correlation function
$C_{L}^{x}(t)\equiv\langle x|\sigma^{x}_{1}(t)\sigma_{L}^{x}(t)|x\rangle-
\langle x|\sigma^{x}_{1}(t)|x\rangle\langle x|\sigma_{L}^{x}(t)|x\rangle$ is simply
\begin{equation}
C_{L}^{x}(t)=|\langle A_{1}B_{L}\rangle_{t}|^{2}\; .
\end{equation}
The simplicity of the formulae obtained so far is due to the
locality in terms of the $A$ and $B$ operators of the non-vanishing expectations entering here. For other quantities,
for example expectation of $\sigma_{l}^{x}$ at a general site $l$, the formula is no more local and one should
use Wick's theorem in order to evaluate the string operators appearing. This leads to huge determinants that
are not efficiently calculable numerically, especially if one has to perform averages over disorder realizations.
In this case, a direct space calculation is probably more appropriate.

\subsection{Results}
In the numerical study, we use the following distribution for the couplings and fields:
\begin{equation}
\pi(J)=\left\{\begin{array}{cl}
1 & \qquad J \in [0,1]\\
0 &\qquad otherwise
\end{array}\right.
\end{equation}
and
\begin{equation}
\rho(h)=\delta(h-h_{o})\; .
\end{equation}
The critical field $h_{c}$ is obtained from the vanishing of the control parameter $\delta$, that is
$h_{c}=e^{-1}$. Average quantities
\begin{equation}
[Q]=\frac{1}{N}\sum_{i=1}^{N}Q^{(i)}
\end{equation}
are computed using typically $N\sim 50000$ disorder realizations for chains of size up to $L\sim 100$.

In figure~1 we show the time dependence of the average surface and bulk magnetisation,
$[m_{z}^{z}(1,t)]$ and $[m_{z}^{z}(L/2,t)]$ respectively, at the critical
point starting with the $|z\rangle$ state. For long times, the non-equilibrium
surface and bulk transverse magnetisations reach a size-independent constant value $m_{z}^{s,b}(\infty)$.
The relaxation toward $m_{z}^{s,b}(\infty)$ is clearly slower than exponential and presumably
algebraic, with superimposed oscillations with period of about $2\pi$. This new time-scale is related to the
disorder distribution properties. We have tested the robustness of those oscillations against disorder distributions.
For exponential like distributions, $P(J)=\alpha\exp(-\alpha J)$, they have a period
$\tau$ varying linearly with the width parameter $\alpha$, meaning that the microscopic time scale $\tau$
is set by the average coupling $[J]$, as $\tau\sim [J]^{-1}=\alpha$. So one expects the presence of these oscillations
for every disorder distributions with a finite first momentum.
\begin{figure}[h]
\epsfxsize=8truecm
\begin{center}
\mbox{\epsfbox{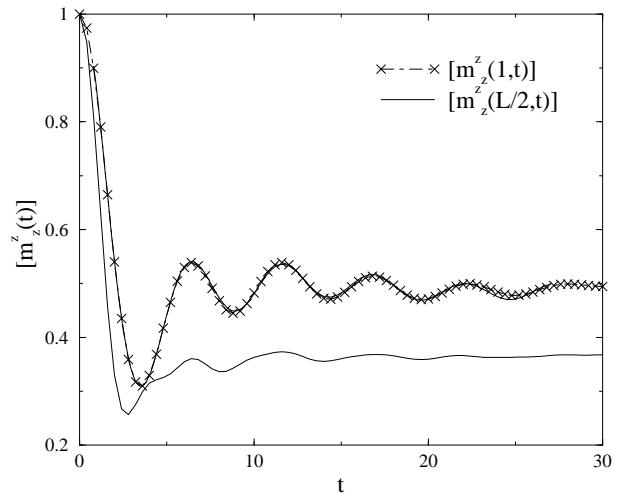}}
\end{center}
\caption{\label{fig1}Surface and bulk non-equilibrium average transverse magnetisation at the critical field
$h_{c}$ starting with the
$|z\rangle$ state. The thick line corresponds to the fit discussed in the text. }
\end{figure}

In the case of the surface magnetisation,
the relaxation behaviour is very well fitted by the form
$t^{-1.15}[\cos(1.2t-1.46)-0.43\cos(0.3t-1)]+0.5$ as shown in figure~1,
which valid the power law relaxation behaviour. For the bulk magnetisation such a fit is much more
hazardous since the oscillations have a more complicated structure.

\begin{figure}[h]
\epsfxsize=8truecm
\begin{center}
\mbox{\epsfbox{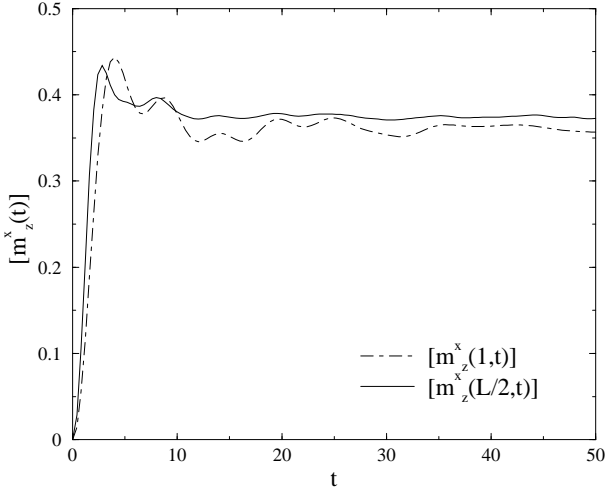}}
\end{center}
\caption{\label{fig2} Surface and bulk non-equilibrium average transverse magnetisation
at the critical field $h_{c}$ starting with
the $|x\rangle$ state.}
\end{figure}

Figure~2 shows the same quantity starting with the $|x\rangle$ state at time $t=0$. The surface and bulk
non-equilibrium transverse magnetisations approache algebraically the constants $m_{x}^{s,b}(\infty)$.
The features observed for the average magnetisation are very similar to the homogeneous critical case where one finds
$m_{z}^{z,x}(t)=1/2\pm J_{1}(4t)/4t$, with the $+$ ($-$) sign for the $z(x)$-state~\cite{trimper,igloi1,karevski}.
Thus the relaxation toward $1/2$, is characterized by the power law $t^{-3/2}$.

The asymptotic values of the non-equilibrium magnetisation are depending on the value of the field $h_{o}$.
We present the dependence of these quantities on the transverse field in figure~3 for increasing sizes for the disordered case.
We give also, in order to compare, the exact asymptotic values for the pure case~\cite{trimper,igloi1}.
The behaviour of the asymptotic averaged magnetisation doesn't seem to have any singularity near the critical
field contrary to the pure case. Nevertheless, finite-size effects should not be excluded too quickly in the
vicinity of the critical point also the curve shown in figure~3 have almost collapsed.
One may notice that as in the homogeneous
situation, the asymptotic limits take the same value at the critical field for both initial states.
\begin{figure}[h]
\epsfxsize=8truecm
\begin{center}
\mbox{\epsfbox{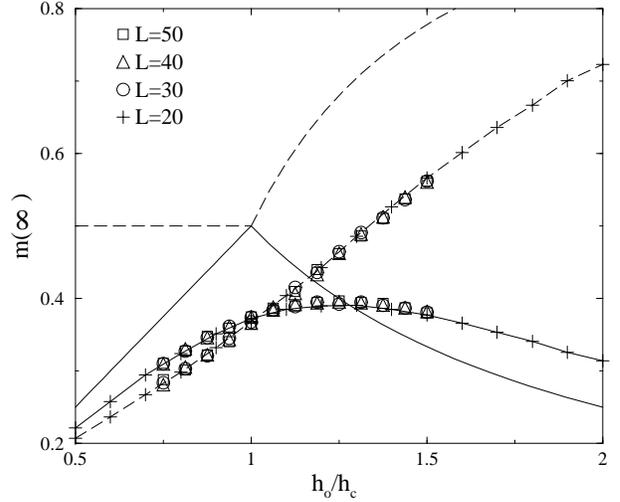}}
\end{center}
\caption{\label{fig3} Asymptotic values of the non-equilibrium average transverse magnetisation
as functions of the distance to the critical field $h_{o}/h_{c}$.  Lines with symbols refer to the disordered
situation while thick lines to the exact pure case solutions. The solid(dashed) line corresponds
to the $|x\rangle$($|z\rangle$).}
\end{figure}

We turn now to the discussion of the average non-equilibrium end-to-end correlation, $[C_{L}^{x,z}(t)]$.
Figure~4 show the behaviour of this quantity for various fields.
\begin{figure}[h]
\epsfxsize=8truecm
\begin{center}
\mbox{\epsfbox{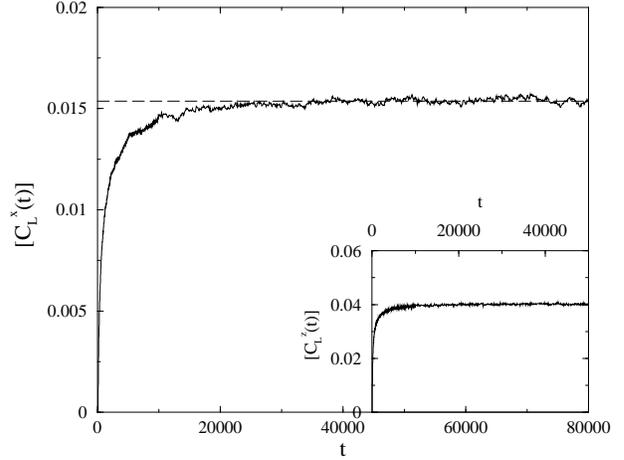}}
\end{center}
\caption{\label{fig4} Average end-to-end correlation functions at the critical field $h_{c}$ and $L=20$
 starting with the
$|x\rangle$ state. The dashed line is given by (\ref{c1}). Insert: same as before for the $|z\rangle$ state. }
\end{figure}
Since the velocity of the excitations are finite, the end-to-end correlations are vanishing up to a threshold time
$\tau(L)\propto L$.
No signal starting from one end has yet reached the other end.
For $t>\tau(L)$, the correlations start to build up and finally for very long
times they reach a finite value
$C_{L}^{x,z}(\infty)$ depending on $L$. In figure~5 we present the size dependence
of the asymptotic value
$C_{L}^{x}(\infty)$ for $h_{o}=h_{c}$, $h_{o}>h_{c}$ and $h_{o}<h_{c}$.
\begin{figure}[h]
\epsfxsize=8truecm
\begin{center}
\mbox{\epsfbox{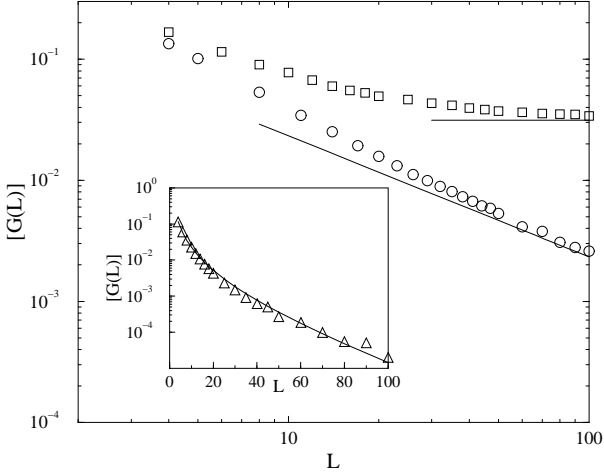}}
\end{center}
\caption{\label{fig5} Size dependence of the asymptotic value $[G(L)]$
for $h_{o}=h_{c}$ (circles), $h_{o}=0.8h_{c}$ (squares). Inset : Size dependence of the asymptotic value
$C_{L}^{x}(\infty)$ for $h_{o}=1.2h_{c}$ (triangles). The solid lines indicate the analytical predictions.}
\end{figure}

The asymptotic behaviour of the end-to-end correlators can be understood as follows. To fix the ideas, consider the simple
connected end-to-end correlator $C_{L}^{x}(t)=|\langle A_{1}B_{L}\rangle_{t}|^{2}$. It can be split into a
time-dependent quantity and a time independent part:
\begin{equation}
C_{L}^{x}(t)=F(L,t)+G(L)\; ,
\end{equation}
where $G(L)=1/2\sum_{q}\phi_{q}^{2}(1)\psi_{q}^{2}(L)$ and
$F(L,t)=\sum_{q\ne q'}\Gamma_{qq'}[\cos(\epsilon_{q}-\epsilon_{q'})t-\cos(\epsilon_{q}+\epsilon_{q'})t]$
with $\Gamma_{qq'}=\phi_{q}(1)\phi_{q'}(1)\psi_{q}(L)\psi_{q'}(L)$. For $t\gg 1$, the arguments of the cosines
are random phases and the $F(L,t)$ contribution vanishes.
In fact, since the bottom spectrum is behaving at the critical point as
$\epsilon_{q}\sim \exp(-const.L^{1/2})$~\cite{igloi2}, the phases will be random for
$t>\epsilon^{-1}\sim \exp(const.L^{1/2})$.
The average correlator
is thus asymptotically given by
\begin{equation}
[C_{L}^{x}(t\gg 1)]\simeq [G(L)]\; .
\end{equation}
In the sum over all the modes, the dominant contribution comes from the lowest mode
which is localized near the surface, the others being exponentially smaller. Thus, for large
sizes one has
\begin{equation}
[G(L)]=\frac{1}{2}\sum_{q}[\phi_{q}^{2}(1)\psi_{q}^{2}(L)]\propto
[\phi_{1}^{2}(1)\psi_{1}^{2}(L)]\; .
\label{c1}
\end{equation}
Physically, for a first excitation $\epsilon_{1}$ vanishing faster than $1/L$,
$\phi_{1}(1)$ and $\psi_{1}(L)$ give the equilibrium magnetisation in the $x$ direction
at both boundaries~\cite{peschel,karevski2}.
One has
\begin{equation}
\langle \sigma_{1}^{x}\rangle=\phi_{1}(1)=
\left(1+\sum_{i=1}^{L-1} \prod_{j=1}^{i} \left( \frac{h_{j}}{J_{j}} \right)^{2} \right)^{-1/2}\; ,
\end{equation}
and a similar expression for $\langle\sigma^{x}_{L}\rangle$ with $h_{j}/J_{j}$ replaced by $h_{L-j}/J_{L-j}$.
At the critical point, for a typical sample the fluctuations of the couplings generated along the chain
lead to
\begin{equation}
\sum_{i}^{L-1}\prod_{j=1}^{i}\left( \frac{h_{j}}{J_{j}} \right)^{2}\sim \exp\left(c L^{1/2}\right)
\end{equation}
thanks to the central limit theorem, where $c$ is a positive constant.
So that the contribution to the average $[G(L)]$ of a typical sample is of order
$\exp(-c L^{1/2})$.
Stronger contributions to the average $[G(L)]$ come from rare samples where the sum
$\sum_{i}^{L-1}\prod_{j=1}^{i} (h_{j}/J_{j})^2$ is of order one. As noticed in several
previous works~\cite{igloi2,karevski3,karevski4}
we can relate this problem to the surviving probability of a one-dimensional walker making $L$ steps with
an absorbing boundary at the origin. The quantity $[G(L)]$ is related to the probability
$P_{L}(0<y_{i}<y_{L})$
that a walker starting near the origin, on the positive side,
ends after $L$ steps at a position $y_{L}$ larger
than all previous positions $y_{i}$ without visiting negative sites. This is due to the fact that in $[G(L)]$
it is the product of both right and left magnetisations that enter.
For $\delta=0$, that is at the critical point, the walker is unbiased
and from the walker interpretation one has $[G(L)]\sim 1/L$.
On the contrary for $\delta>0$ there is a drift
proportional to $\delta$ toward the absorbing boundary.
In this case, contributions of order one are very rare
and the average $[G(L)]$, proportional to the surviving probability,
scales as $L^{-3/2}\exp(-aL)$ with $a>0$ a constant~\cite{igloi2}.
This form fits very well the datas as it could be seen
in the inset of figure~5.
For $\delta<0$,
that is in the ferromagnetic phase, the walker
is drifted off the absorbing boundary and for $L$ sufficiently large, the typical fluctuations of order $L^{1/2}$ are
not sufficient to make the walker cross the origin. $[G(L)]$ is reaching a finite constant for
$L$ large enough.
It is interesting to notice again here that it is the connected part of the average end-to-end non-equilibrium correlator
$[C_{L}^{x}(t)]$ that reaches a constant for $t\gg 1$.
These three different regimes are summarized in figure~5.
Same arguments can be applied to explain the similar behaviour of the average correlator $[C_{L}^{z}(t)]$.

\section{Summary}
We have studied numerically the non-equilibrium relaxation properties of the random Ising quantum chain
by calculating the average transverse magnetisation and end-to-end correlation functions. Starting with a
fully magnetised initial state, the random system relaxes toward a stationary state. The relaxation of the
average transverse magnetisation, toward a constant value depending on the transverse field, is clearly not
exponential due to the presence of too large oscillations, but most probably algebraic.
For the same reasons it is hard to extract from the numerics the exact algebraic decay. This behaviour looks
like the decay of the pure model without any disorder. Nevertheless, contrary to the pure case where quantum interference
plays a central role, in the disordered system the averaging process over many samples is
predominant.

The time-evolution of the end-to-end correlations, which gives some simple insight of the development of the
spatial correlations in the quantum chain, is somehow more instructive. For times smaller than a threshold
time $\tau(L)$, they vanish since both ends are not causally connected. After this limiting time, they start
to grow up and for very long times they finally reach a constant value depending on the transverse field and on the
system size. For the $|x\rangle$ initial state, utilising a random walk argument, one can explain the scaling
of this stationary size-dependent value. At the critical field, the main contribution to the average correlation
comes from rare samples that are strongly correlated due to large domains of strong bonds,
giving a contribution of order one,
while the contribution of a typical sample is exponentially small with the system size. So the average value
is completely determined by the distribution of rare samples. This leads to a $1/L$ scaling behaviour of the
stationary value. In the ordered phase, the contribution of strongly correlated domains is very important and
gives rise to
a size-independent stationary value, as confirmed by the drifted walker analogy. On contrary, for a large
transverse field, the walker is driven toward the absorbing boundary, as explained in the previous section, and
the stationary value behaves as $\exp(-c L)$. In conclusion, contrary to the pure chain where
after a sudden jump to a value $L^{-a}$ at $t=\tau(L)$, the threshold
time~\cite{igloi1}, the end-to-end correlation  decreases toward zero with a stretched exponential like behaviour,
in the disordered case the asymptotic behaviour is simpler. This discrepancy lies again in the fact that
quantum interferences in the pure
case are important while they play no significant role in the averaging process in the disordered situation.
The stationary behaviour is just related to the statistical distribution of large strongly correlated domains.

Acknowledgements: Useful discussions with Christophe Chatelain are greatefully acknowledged.

\end{document}